\documentclass[english,prl,aps,twocolumn,groupedaddress,nofootinbib,superscriptaddress,longbibliography,floatfix]{revtex4-1}
\usepackage{amsmath,amssymb,bm,graphicx,booktabs,array}
\usepackage[colorlinks=true,linkcolor=blue,citecolor=blue,urlcolor=blue]{hyperref}
\newcommand{\avg}[1]{\langle #1 \rangle}
\begin{document}

\title{Precision Hypernuclear Structure from Interaction Cross Sections}
\author{Carlos A. Bertulani}
\email{carlos.bertulani@etamu.edu}
\affiliation{Department of Physics, East Texas A\&M University, Commerce, Texas 75429, USA}
\affiliation{Technische Universit\"at Darmstadt, Institut f\"ur Kernphysik, 64289 Darmstadt, Germany}
\affiliation{ExtreMe Matter Institute EMMI, GSI Helmholtzzentrum f\"ur Schwerionenforschung GmbH, D-64291 Darmstadt, Germany}

\begin{abstract}
We establish interaction cross sections as a quantitative structural observable for hypernuclei. Using the hypertriton, ${}^{3}_{\Lambda}\mathrm H$, as the most stringent benchmark, microscopic three-body wave functions are combined with a finite-range coupled-channel Glauber calculation and Bayesian inversion. For ${}^{3}_{\Lambda}\mathrm{H}+{}^{12}\mathrm{C}$ at 1.5 GeV/nucleon, the interaction cross section decreases from about 1150 to 750 mb as the $\Lambda$ separation energy $B_\Lambda$ increases from 0.1 to 0.5 MeV, while propagated reaction-model uncertainties remain near 5\%. A measurement therefore supplies a direct inversion $\sigma_I\to r_m\to B_\Lambda$, constraining both the matter radius and the $\Lambda$ separation energy. The framework provides the missing quantitative link between proposed interaction-cross-section measurements and the spatial structure of an already formed hypernucleus. Although demonstrated for the hypertriton, the methodology is readily extendable to other weakly bound hypernuclei.
\end{abstract}
\maketitle

\textit{Introduction.} -- Determining the spatial structure of a bound hypernucleus remains difficult. Spectroscopy constrains level energies, weak decays probe decay dynamics, and production or femtoscopy observables depend on reaction evolution, freeze-out, coalescence, or source modeling. None of these approaches directly determines the matter distribution of an already formed hypernucleus through a reaction observable with quantified theoretical uncertainties. A dedicated FAIR program has therefore proposed interaction-cross-section measurements of the hypertriton, and a two-target strategy has been developed to reveal a hypernuclear halo \cite{GSIproposal,Velardita2023}. The missing ingredient is a quantitative map from a measured interaction cross section to the matter radius and separation energy. This letter provides that map.

The experimental motivation is timely. Modern measurements have substantially improved our knowledge of the hypertriton lifetime and production systematics, while determinations of its $\Lambda$ separation energy, $B_\Lambda$, still exhibit a significant spread. A radius measurement would provide an independent constraint on the long-distance wave function and on the low-energy $\Lambda d$ dynamics that underlies halo-EFT descriptions. Interaction cross sections are uniquely suited to this purpose because they are inclusive, geometrical observables with a long history of matter-radius extraction in exotic nuclei \cite{Tanihata1985,Ozawa2001}.

The hypertriton is the lightest known hypernucleus and one of the most weakly bound hadronic systems. Its $\Lambda$ separation energy, $B_\Lambda\sim0.1$--$0.5$ MeV, implies a spatially extended $\Lambda+d$ configuration and a hyperon-core separation approaching 10 fm \cite{Juric1973,Davis2005,Ji2012,Hammer2017,HIHa19}. It is consequently a benchmark for hyperon-nucleon interactions, few-body universality, and the long-standing hypertriton puzzle \cite{STAR2020,ALICE2022,HypHI2013,HADES2022,ALICE2026}. The near-threshold character amplifies structural sensitivity because the halo size grows approximately as $r_m\propto B_\Lambda^{-1/2}$.

We show that this universal amplification turns the interaction cross section into a precision structural observable. The result is not merely a prediction for one reaction: it establishes a general experimental methodology in which a reaction measurement is inverted into hypernuclear structure,
\begin{equation}
\sigma_I\longrightarrow r_m\longrightarrow B_\Lambda .
\label{eq:map}
\end{equation}
This capability is qualitatively different from spectroscopy, lifetime, production, and femtoscopy measurements because it directly constrains the spatial wave function of a formed hypernucleus.

\textit{Three-body hypertriton structure.} -- The hypertriton is described as $p+n+\Lambda$ in Jacobi coordinates using a stochastic variational expansion in correlated Gaussians \cite{Suzuki1998,Hiyama2003}. The dominant deuteron-like $pn$ configuration has $S_{pn}=1$, $T_{pn}=0$, and predominantly $l_x=0$, coupled to the $\Lambda$ in an $l_y=0$ relative state. Deuteron $D$-wave components and correlated $x$--$y$ Gaussian terms are included so that both short-range correlations and the very extended $\Lambda+d$ tail are represented. An AV18-inspired Gaussian representation is used for the $NN$ interaction, while a practical spin-dependent $YN$ interaction is varied to generate $0.10\le B_\Lambda\le0.50$ MeV. The basis exponents are distributed over a broad logarithmic range so that the same expansion resolves both the deuteron scale and the much longer $\Lambda d$ scale. The asymptotic tail is constrained by the binding momentum associated with the chosen $B_\Lambda$. This point is essential: assigning the long hyperon tail indiscriminately to all particle coordinates would spuriously enlarge the proton and neutron radii. The one-body densities used in the reaction calculation are instead constructed with exact center-of-mass transformations from the Jacobi coordinates.

The calculated matter radii are broadly consistent with few-body and effective-field-theory studies \cite{Ji2012,Hammer2017,HIHa19}. Hildenbrand and Hammer found the rapid growth of the hypertriton radius as $B_\Lambda$ approaches zero, while Ji, Phillips, and Platter showed that the spatial extent is governed primarily by the small $\Lambda$ separation energy. Here $r_m$ ranges from about 2.8 fm at $B_\Lambda=0.50$ MeV to 5.7 fm at 0.10 MeV. The ratio $r_m(0.10)/r_m(0.50)\simeq2.1$ is close to the near-threshold estimate $\sqrt{0.50/0.10}=2.24$ expected from the leading halo scaling $r\propto B_\Lambda^{-1/2}$ \cite{Ji2012,Hammer2017,HIHa19}, supporting the physical consistency of the halo tail.

The calibration also has a residual dependence on the microscopic $YN$ interaction that is distinct from the reaction-model uncertainty propagated below. At fixed \(B_\Lambda\), the asymptotic \(\Lambda d\) binding momentum fixes the leading long-distance behavior. Since both \(r_m\) and \(\sigma_I\) are particularly sensitive to this halo tail, their correlation is expected to be less model dependent than short-distance observables. Different realistic $YN$ interactions or EFT regularizations can nevertheless modify effective-range corrections, short-range correlations, and channel admixtures, producing a residual shift of the $\sigma_I$ - $r_m$ - $B_\Lambda$ calibration. A systematic multi-Hamiltonian analysis will therefore be required once experimental precision warrants including this structure-model uncertainty.

\begin{table}[t]
\centering
\caption{Representative validation wave function at $B_\Lambda=0.133$ MeV. Jacobi-coordinate radii differ from one-body center-of-mass radii.}
\label{tab:wf}
\begin{tabular}{lc}
\hline\hline
Quantity & Value \\
\hline
$E_{\rm total}$ & $-2.357$ MeV \\
$B_{\Lambda}^{\rm model}$ & $0.1329$ MeV \\
One-body matter rms radius & $5.253$ fm \\
One-body hyperon rms radius & $6.973$ fm \\
$r_{\Lambda-NN}$ & $10.34$ fm \\
$r_{NN}$ & $2.960$ fm \\
\hline\hline
\end{tabular}
\end{table}

Table~\ref{tab:wf} lists a representative validation point. The complete wave-function diagnostics and definitions are given in the Supplemental Material \cite{supplemental}.

\textit{Coupled-channel reaction framework.} -- The reaction calculation uses the finite-range optical-limit Glauber approximation, folding the microscopic proton, neutron, and hyperon densities with the ${}^{12}$C target density. The elementary $NN$, $\Lambda N$, $\Sigma N$, and transition $\Lambda N\leftrightarrow\Sigma N$ profile functions are constrained by available high-energy scattering systematics. Their cross sections, real-to-imaginary ratios, and finite-range slopes, together with target-density uncertainties, are sampled in a Monte Carlo propagation. The uncertainty of the comparatively poorly known hyperon-target interaction is therefore explicitly included in the bands shown below. Schematically, each opacity is obtained from
\begin{equation}
\Omega_{ab}(\mathbf b,\xi)=\sum_{j\in P}\int d^2s\,T_T(\mathbf s)\,
\Gamma^{(jT)}_{ab}(\mathbf b+\mathbf s_j-\mathbf s),
\label{eq:opacity}
\end{equation}
where $a,b\in\{\Lambda,\Sigma\}$, $T_T$ is the target thickness, and $\Gamma_{ab}$ is the finite-range elementary profile. Nucleon-target scattering is diagonal in hyperon space, whereas the hyperon terms generate both diagonal attenuation and transition opacities. The calculation is therefore an optical-limit multiple-scattering theory, but it retains the full projectile configuration dependence required for diffractive breakup.

Propagation occurs in the coupled hyperon basis $\{|\Lambda\rangle,|\Sigma\rangle\}$. For a projectile configuration $\xi=\{\mathbf s_p,\mathbf s_n,\mathbf s_\Lambda\}$ and impact parameter $\mathbf b$,
\begin{equation}
\hat S(\mathbf b,\xi)=\exp[-\hat\Omega(\mathbf b,\xi)/2],\quad
\hat\Omega=
\begin{pmatrix}
\Omega_{\Lambda\Lambda}&\Omega_{\Lambda\Sigma}\\
\Omega_{\Sigma\Lambda}&\Omega_{\Sigma\Sigma}
\end{pmatrix}.
\label{eq:Smat}
\end{equation}
The diagonal opacities describe propagation within a hyperon channel, whereas the off-diagonal elements transfer flux through $\Lambda N\leftrightarrow\Sigma N$. The $\Lambda$ survival amplitude $S_{\Lambda\Lambda}=[e^{-\hat\Omega/2}]_{\Lambda\Lambda}$ therefore contains virtual $\Lambda\to\Sigma\to\Lambda$ excursions to all orders. For an incoming state $|\Psi_0,\Lambda\rangle$,
\begin{equation}
P_{\rm el}(b)=|\langle\Psi_0,\Lambda|\hat S|\Psi_0,\Lambda\rangle|^2,
\qquad
\sigma_I=\int d^2b\,[1-P_{\rm el}(b)].
\label{eq:sigmaI}
\end{equation}

The interaction loss is separated into diffractive breakup, explicit $\Lambda\to\Sigma$ conversion, and absorption. The Good-Walker fluctuation term is \cite{GW60,Bertulani2004,Glauber1959,Bertulani2004a}
\begin{equation}
\sigma_{\rm breakup}=\int d^2b\left[\avg{|S_{\Lambda\Lambda}|^2}_0-|\avg{S_{\Lambda\Lambda}}_0|^2\right],
\label{eq:break}
\end{equation}
while
\begin{equation}
\sigma_{\Lambda\Sigma}=\int d^2b\,\avg{|S_{\Sigma\Lambda}|^2}_0,
\qquad
\sigma_{\rm abs}=\sigma_I-\sigma_{\rm breakup}-\sigma_{\Lambda\Sigma}.
\label{eq:channels}
\end{equation}
The three terms are exclusive by construction. Equation~(\ref{eq:break}) is the variance of the elastic amplitude over projectile configurations: it vanishes for a structureless projectile but grows when compact and extended halo configurations encounter substantially different opacities. Equation~(\ref{eq:channels}) identifies explicit hyperon conversion through the off-diagonal $S$-matrix element; all remaining loss, including stripping and inelastic removal of constituents, is assigned to absorption. This organization prevents breakup or conversion from being counted twice. Further details, including the opacity folding and analytic $2\times2$ matrix exponential, are given in the Supplemental Material  \cite{supplemental}.

\begin{figure}[t]
\centering
\includegraphics[width=0.48\textwidth]{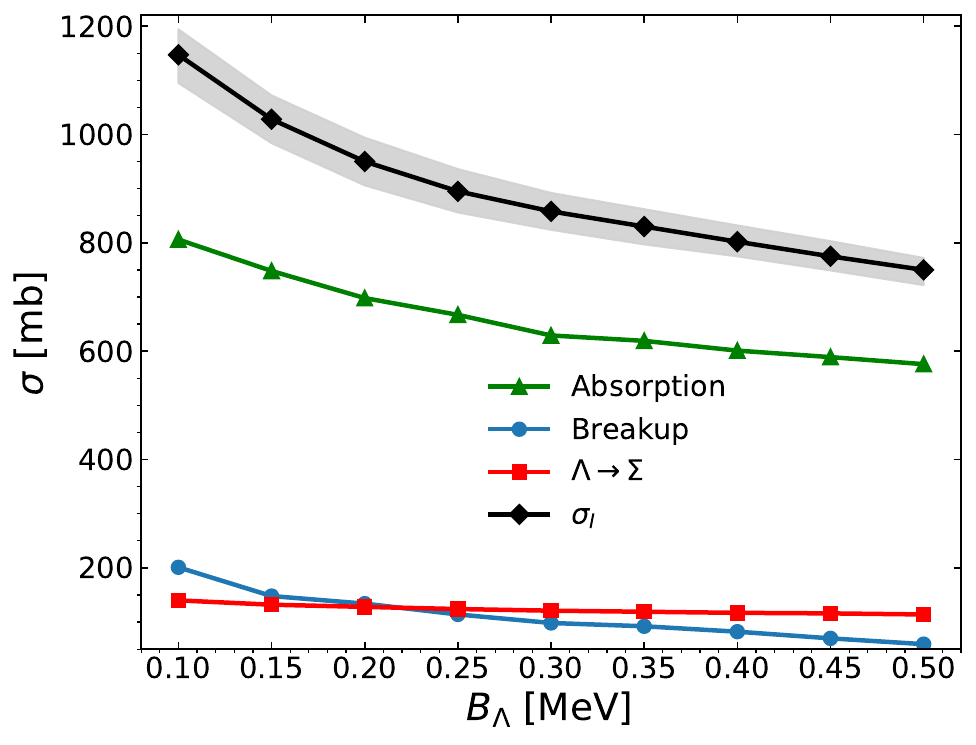}
\caption{Channel decomposition, $\sigma_I=\sigma_{\rm breakup}+\sigma_{\Lambda\Sigma}+\sigma_{\rm abs}$, for ${}^{3}_{\Lambda}\mathrm H+{}^{12}\mathrm C$ at 1.5 GeV/nucleon. The gray band is the propagated one-standard-deviation interval.}
\label{fig:sigmaBL}
\end{figure}

Figure~\ref{fig:sigmaBL} displays the central result. Increasing $B_\Lambda$ from 0.10 to 0.50 MeV contracts the halo and lowers $\sigma_I$ from about 1150 to 750 mb. This 400-mb lever arm is about an order of magnitude larger than the propagated one-standard-deviation uncertainty. The breakup contribution decreases most rapidly with binding because it is generated by fluctuations among extended and compact configurations. The $\Lambda\Sigma$ term varies more weakly, while absorption remains the largest contribution.

The covariance analysis shows that the uncertainty of $\sigma_I$ is dominated by the absorptive hyperon-nucleus sector, which accounts for about 80--90\% of the propagated variance. The $\Lambda\Sigma$ channel contributes typically 10--20\%, while breakup has only a small net covariance share. These percentages describe uncertainty propagation, not the relative central strengths of the channels. The resulting total uncertainty remains close to or below 5\% over the full interval. Correlations among channels are important: breakup can have a sizable individual spread while contributing little to the variance of the sum because its fluctuations anticorrelate with absorption. The full covariance matrices are therefore reported in the Supplemental Material rather than estimating the total error by adding independent channel widths in quadrature \cite{supplemental}. Unlike production observables, $\sigma_I$ probes the spatial distribution after the hypernucleus has formed and is therefore largely decoupled from transport, freeze-out, and coalescence assumptions. This makes the observable complementary rather than redundant: it tests the long-distance structure directly, whereas production and lifetime measurements test formation and decay dynamics.

\begin{figure}[t]
\centering
\includegraphics[width=0.48\textwidth]{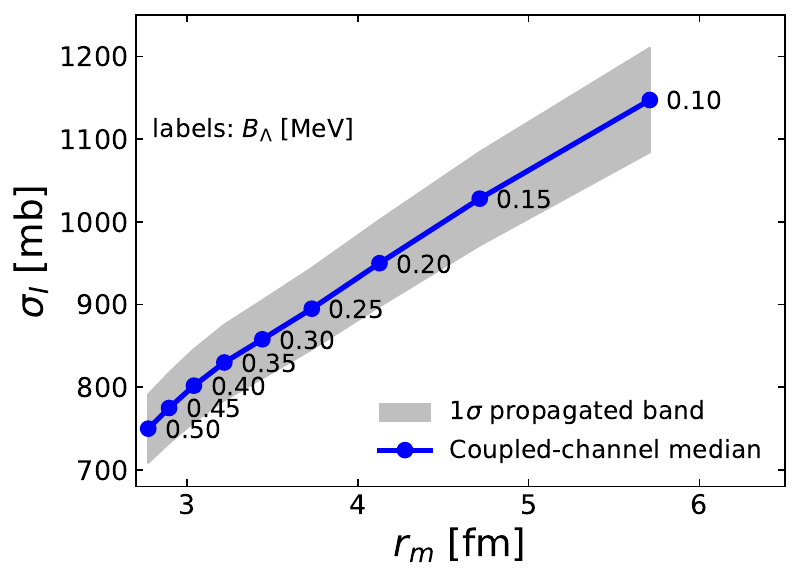}
\caption{Interaction cross section versus the calculated hypertriton matter radius. Labels give $B_\Lambda$ in MeV.}
\label{fig:sigmarm}
\end{figure}

\textit{Matter-radius calibration.} -- The same results plotted against $r_m$ in Fig.~\ref{fig:sigmarm} give a nearly one-to-one calibration curve. Over most of the interval, a linear fit yields
\begin{equation}
\frac{d\sigma_I}{dr_m}\simeq130~\mathrm{mb/fm},
\end{equation}
corresponding to an approximately 13\% change in the interaction cross section per femtometer. \begin{table}[t]
\centering
\caption{Calibration values connecting $r_m$, $B_\Lambda$, and $\sigma_I$.}
\label{tab:calibration}
\resizebox{\columnwidth}{!}{%
\begin{tabular}{lccccccccc}
\hline\hline
$r_m$ (fm)&5.71&4.71&4.13&3.73&3.44&3.22&3.04&2.89&2.77\\
\hline
$B_\Lambda$ (MeV)&0.10&0.15&0.20&0.25&0.30&0.35&0.40&0.45&0.50\\
$\sigma_I$ (mb)&1147&1028&950&895&858&830&802&775&750\\
\hline\hline
\end{tabular}}
\end{table}

Interaction cross sections have transformed matter-radius measurements for ordinary exotic nuclei \cite{Tanihata1985,Ozawa2001,Teixeira2021}; here the sensitivity is stronger because the hypertriton occupies an extreme halo regime. A measured $\sigma_I$ can therefore be converted directly into $r_m$, with the complete calibration values tabulated in Table~\ref{tab:calibration}. The relation is monotonic over the full interval, so the inverse problem is not affected by multiple radius solutions. At the weak-binding end the curve becomes steeper because a small change in $B_\Lambda$ strongly modifies the halo tail; at larger binding the system becomes more compact and the response gradually saturates. This is precisely the behavior expected for a near-threshold few-body state and provides a reaction-level manifestation of halo universality.

\begin{figure}[t]
\centering
\includegraphics[width=0.48\textwidth]{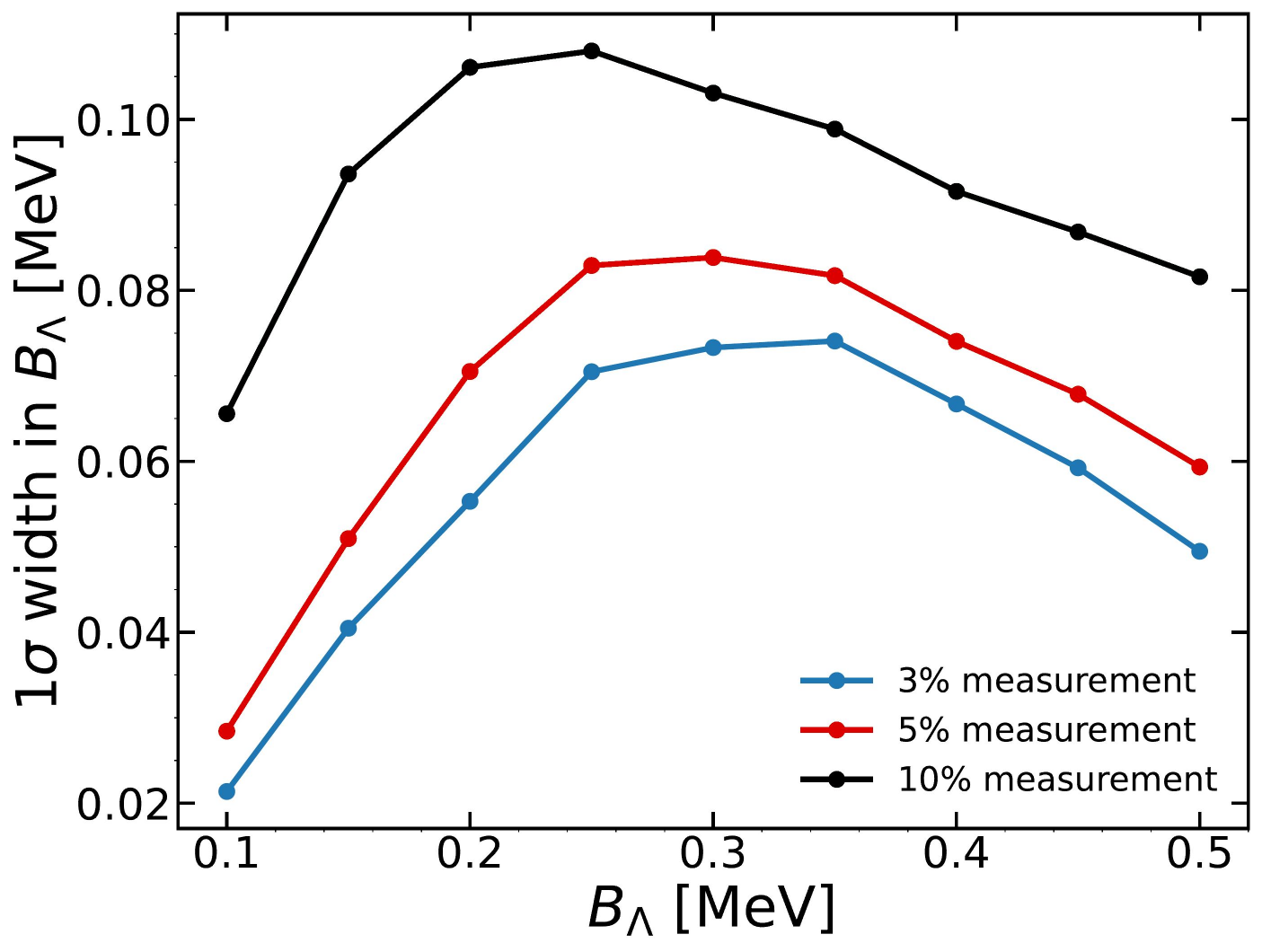}
\caption{Forecast one-standard-deviation width in $B_\Lambda$ for assumed 3\%, 5\%, and 10\% relative uncertainties on the measured $\sigma_I$, using a uniform prior over $0.1\leq B_\Lambda\leq0.5$ MeV.}
\label{fig:bayesian}
\end{figure}

\textit{Bayesian inversion.} -- For a measured $\sigma_I$, the likelihood combines experimental and propagated theoretical uncertainties. A uniform prior over $0.1\le B_\Lambda\le0.5$ MeV is used. For a hypothetical measurement $\sigma_I^{\rm exp}$, the likelihood is
\begin{equation}
p(D|B_\Lambda)\propto\exp\left[-\frac{(\sigma_I^{\rm exp}-\sigma_I^{\rm th}(B_\Lambda))^2}{2(\Delta_{\rm exp}^2+\Delta_{\rm th}^2)}\right],
\label{eq:likelihood}
\end{equation}
where the theoretical width is interpolated from the Monte Carlo covariance. 

For multiple measurements with correlated systematic uncertainties, the likelihood is generalized by replacing the scalar variance in Eq. (\ref{eq:likelihood}) with the full experimental-plus-theoretical covariance matrix. Common normalization, target-thickness, efficiency, or other systematic uncertainties can equivalently be introduced as nuisance parameters and marginalized in the posterior.

Figure~\ref{fig:bayesian} shows that a representative measurement with 5\% relative uncertainty gives $\delta B_\Lambda\simeq0.05$--$0.10$ MeV and a matter-radius uncertainty of only a few tenths of a femtometer. Within the reaction-model uncertainties quantified here, the sensitivity is sufficiently large that experimental precision becomes the dominant limitation.

The broad maximum in Fig.~\ref{fig:bayesian} has both physical and statistical origins. The slope $|d\sigma_I/dB_\Lambda|$ is largest at weak binding, where the halo changes rapidly, and decreases as the system becomes compact. In addition, the finite prior truncates posterior distributions near 0.1 and 0.5 MeV, reducing the corresponding credible widths. Extending the prior to a broader interval raises the endpoint uncertainties and flattens the maximum, but it does not remove the large information content set locally by $\delta B_\Lambda\sim\delta\sigma_I/|d\sigma_I/dB_\Lambda|$.

For the light ${}^{12}$C target, nuclear interactions dominate. A separate finite calculation gives a pure Coulomb-breakup contribution of 8.6 mb at 1.5 GeV/nucleon, below the quoted reaction-model uncertainty, and it is therefore not included in $\sigma_I$. This treatment avoids inserting an unregularized long-range Coulomb eikonal phase, which is known to produce a divergent breakup amplitude. Heavy targets would require an explicit finite Coulomb-breakup treatment and are outside the present scope.

\textit{Experimental prospects and broader implications.}---The proposed FAIR measurement would provide the first direct test of the calibration in Eq.~(\ref{eq:map}). Such measurements remain challenging because of low production rates and the need for efficient weak-decay vertex reconstruction. A dedicated sensitivity study must ultimately include beam intensity, target configuration, acceptance, backgrounds, and reconstruction efficiency. The present result does not assume that a particular precision will be reached; rather, it demonstrates that even 
a measurement with 5--10\% relative uncertainty would provide meaningful structural information because the physical variation is hundreds of millibarns and the reaction-theory uncertainty is smaller.

A rough counting estimate indicates that a 10\% statistical determination would require samples of order $10^4$ reconstructed hypertritons, although the actual requirement depends strongly on transmission geometry, decay losses, detector acceptance, and background subtraction. This number should therefore be regarded as a scale estimate rather than an experimental forecast. The important result here is that a moderate-precision measurement already maps onto a useful interval in both $r_m$ and $B_\Lambda$.

Although the hypertriton is the cleanest and most demanding benchmark, neither the density-folding framework nor the coupled-channel propagation is specific to $A=3$. The same strategy can be applied to heavier single-\(\Lambda\) hypernuclei, neutron-rich hypernuclei, and future hyperhalo candidates by replacing the projectile densities and enlarging the channel space where necessary. The broader advance is therefore methodological: interaction cross sections supply a direct quantitative probe of the spatial wave function of an already formed hypernucleus. For heavier systems, the same framework can constrain neutron-rich cores and hyperon spatial distributions separately when combined with charge-changing or two-target measurements. Thus the present demonstration links the mature interaction-cross-section program of rare-isotope physics to a new class of observables in hypernuclear physics.

For heavier hypernuclei, the dominant theoretical limitation is expected to shift increasingly toward the microscopic many-body proton, neutron, and hyperon densities. Depending on the system and excitation thresholds, an enlarged coupled-channel space may also be required, while uncertainties in the elementary hyperon-nucleon amplitudes and their propagation into effective hyperon-nucleus interactions will remain important.

\textit{Conclusions.}---We have established a quantitative inversion from an interaction cross section to the matter radius and separation energy of a bound hypernucleus. For the hypertriton, universal near-threshold expansion produces a 400-mb lever arm in $\sigma_I$, while a finite-range coupled-channel Glauber treatment keeps propagated uncertainties near 5\%. The result provides a new and independent constraint on the hypertriton puzzle and a general reaction methodology for precision hypernuclear structure.

\begin{acknowledgments}
I acknowledge useful discussions with T. Aumann, H.-W. Hammer, and A. Obertelli. This work was supported by the U.S. Department of Energy under Grant No. DE-SC0026074 and by the ExtreMe Matter Institute EMMI at the GSI Helmholtzzentrum f\"ur Schwerionenforschung.
\end{acknowledgments}

{\bf Supplemental Material}

\textit{Three-body wave functions.} -- The hypertriton is treated as $p+n+\Lambda$ in Jacobi coordinates $\mathbf x=\mathbf r_p-\mathbf r_n$ and $\mathbf y=\mathbf r_\Lambda-(\mathbf r_p+\mathbf r_n)/2$. The variational wave function is expanded in correlated Gaussians over a broad logarithmic range of length scales. The basis includes the dominant deuteron-like $pn$ $S$ wave, a deuteron $D$-wave component, and correlated $x$--$y$ terms. The $YN$ strength is varied to span $0.10\le B_\Lambda\le0.50$ MeV while preserving the short-distance structure. Table~\ref{tab:wf} gives the validation point used in the main text.

\begin{table}[h]
\centering
\caption{Representative hypertriton wave-function properties for $B_\Lambda=0.133$ MeV. Jacobi-coordinate radii are distinct from one-body center-of-mass radii.}
\label{tab:wf}
\begin{tabular}{lc}
\hline\hline
Quantity & Value \\
\hline
$E_{\rm total}$ & $-2.357$ MeV \\
$E_d^{\rm model}$ & $-2.224$ MeV \\
$B_{\Lambda}^{\rm model}$ & $0.1329$ MeV \\
$B_{\Lambda}^{\rm tail}$ & $0.1300$ MeV \\
One-body proton rms radius & $4.1328$ fm \\
One-body neutron rms radius & $4.132$ fm \\
One-body hyperon rms radius & $6.973$ fm \\
One-body matter rms radius & $5.253$ fm \\
$r_{\Lambda-NN}$ & $10.34$ fm \\
$r_{NN}$ & $2.960$ fm \\
$r_{N-\Lambda N}$ & $3.917$ fm \\
$r_{\rm geo}$ & $4.498$ fm \\
\hline\hline
\end{tabular}
\end{table}

The radii increase from about 2.8 fm at $B_\Lambda=0.50$ MeV to 5.7 fm at 0.10 MeV. Their ratio, $r_m(0.10)/r_m(0.50)\simeq2.1$, is close to the universal weak-binding estimate $\sqrt{0.50/0.10}=2.24$.

\textit{Finite-range optical-limit Glauber model.} -- The projectile one-body densities are folded with the ${}^{12}$C density and finite-range elementary profile functions. Schematically,
\begin{equation}
\Omega_{ab}(\mathbf b,\xi)=\sum_{j\in P}\int d^2s\,T_T(\mathbf s)\,\Gamma^{(jT)}_{ab}(\mathbf b+\mathbf s_j-\mathbf s),
\end{equation}
where $a,b\in\{\Lambda,\Sigma\}$ and the nucleon contributions are diagonal in hyperon space. The elementary profiles are parameterized by total cross sections, real-to-imaginary ratios, and finite-range slopes. The central calculation uses high-energy $NN$, $\Lambda N$, and $\Sigma N$ systematics at 1.5 GeV/nucleon; the $\Lambda N\leftrightarrow\Sigma N$ transition strength is varied independently. The Monte Carlo bands sample the elementary cross sections, profile slopes, real parts, target density, and finite-range parameters. This is the optical-limit approximation referred to in the main text.

The coupled-channel operator is
\begin{equation}
\hat S=e^{-\hat\Omega/2},\qquad
\hat\Omega=\begin{pmatrix}
\Omega_{\Lambda\Lambda}&\Omega_{\Lambda\Sigma}\\
\Omega_{\Sigma\Lambda}&\Omega_{\Sigma\Sigma}
\end{pmatrix}.
\end{equation}
For a real symmetric $2\times2$ matrix, defining
\begin{equation}
\bar\Omega=\frac{\Omega_{\Lambda\Lambda}+\Omega_{\Sigma\Sigma}}{2},\quad
\Delta=\frac{\Omega_{\Lambda\Lambda}-\Omega_{\Sigma\Sigma}}{2},\quad
q=\sqrt{\Delta^2+\Omega_{\Lambda\Sigma}^2},
\end{equation}
one obtains
\begin{align}
S_{\Lambda\Lambda}&=e^{-\bar\Omega/2}\left[\cosh\frac q2-\frac{\Delta}{q}\sinh\frac q2\right],\\
S_{\Sigma\Lambda}&=-e^{-\bar\Omega/2}\frac{\Omega_{\Lambda\Sigma}}{q}\sinh\frac q2.
\end{align}
Hence the survival and conversion probabilities depend on all opacity-matrix elements, not only on $\Omega_{\Lambda\Lambda}$.

\textit{Channel decomposition.} -- For the incoming state $|\Psi_0,\Lambda\rangle$,
\begin{equation}
P_{\rm el}(b)=|\langle\Psi_0,\Lambda|\hat S|\Psi_0,\Lambda\rangle|^2,
\qquad \sigma_I=\int d^2b\,[1-P_{\rm el}(b)].
\end{equation}
The Good--Walker diffractive breakup term is
\begin{equation}
\sigma_{\rm breakup}=\int d^2b\left[\avg{|S_{\Lambda\Lambda}|^2}_0-|\avg{S_{\Lambda\Lambda}}_0|^2\right],
\end{equation}
and explicit conversion is
\begin{equation}
\sigma_{\Lambda\Sigma}=\int d^2b\,\avg{|S_{\Sigma\Lambda}|^2}_0.
\end{equation}
The residual loss defines
\begin{equation}
\sigma_{\rm abs}=\sigma_I-\sigma_{\rm breakup}-\sigma_{\Lambda\Sigma},
\end{equation}
so the three terms are exclusive by construction and satisfy the exact sum rule used in the main text.

\begin{table}[h]
\centering
\caption{Channel decomposition. All cross sections are in mb.}
\label{tab:channels}
\begin{tabular}{ccccc}
\hline\hline
$B_\Lambda$ (MeV) & $\sigma_{\rm breakup}$ & $\sigma_{\Lambda\Sigma}$ & $\sigma_{\rm abs}$ & $\sigma_I$ \\
\hline
0.10 & 201 & 140 & 806 & $1147^{+47}_{-52}$ \\
0.15 & 148 & 132 & 748 & $1028^{+44}_{-44}$ \\
0.20 & 134 & 128 & 698 & $950^{+44}_{-44}$ \\
0.25 & 114 & 124 & 667 & $895^{+41}_{-39}$ \\
0.30 & 98.3 & 121 & 629 & $858^{+34}_{-34}$ \\
0.35 & 92.2 & 119 & 619 & $830^{+32}_{-33}$ \\
0.40 & 82.2 & 117 & 601 & $802^{+30}_{-27}$ \\
0.45 & 69.9 & 116 & 589 & $775^{+28}_{-26}$ \\
0.50 & 59.2 & 114 & 576 & $750^{+22}_{-28}$ \\
\hline\hline
\end{tabular}
\end{table}

\begin{table}[h]
\centering
\caption{Calibration values used for $\sigma_I(r_m)$.}
\begin{tabular}{lccccccccc}
\hline\hline
$r_m$ (fm) & 5.71&4.71&4.13&3.73&3.44&3.22&3.04&2.89&2.77\\
\hline
$B_\Lambda$ (MeV)&0.10&0.15&0.20&0.25&0.30&0.35&0.40&0.45&0.50\\
$\sigma_I$ (mb)&1147&1028&950&895&858&830&802&775&750\\
\hline\hline
\end{tabular}
\end{table}

\textit{Uncertainty propagation and covariance.} -- The total uncertainty is propagated by Monte Carlo sampling of $NN$, $\Lambda N$, $\Sigma N$, $\Lambda N\leftrightarrow\Sigma N$, density, and finite-range/profile inputs. The covariance share of channel $i$ is defined by $100\,\mathrm{Cov}(\sigma_i,\sigma_I)/\mathrm{Var}(\sigma_I)$. This quantity can be negative because channel fluctuations are correlated and the channel sum is constrained.

\begin{table}[h]
\centering
\caption{Monte Carlo uncertainty budget. Widths are in mb.}
\small
\begin{tabular}{cccccccc}
\hline\hline
$B_\Lambda$ & $\Delta\sigma_I$ & $\Delta\sigma_{\rm breakup}$ & share & $\Delta\sigma_{\Lambda\Sigma}$ & share & $\Delta\sigma_{\rm abs}$ & share\\
(MeV)&(mb)&(mb)&(\%)&(mb)&(\%)&(mb)&(\%)\\
\hline
0.10&50.1&56.0&-1.8&37.7&11.9&74.7&89.9\\
0.15&47.3&43.6&-0.7&35.0&15.9&63.7&84.9\\
0.20&43.1&32.7&0.1&33.6&13.2&57.0&86.7\\
0.25&38.4&28.7&-0.3&32.1&19.3&50.8&80.9\\
0.30&33.2&26.1&0.5&32.6&12.9&48.7&86.6\\
0.35&32.0&20.5&2.0&31.5&15.0&43.3&83.0\\
0.40&28.2&19.7&-1.4&30.6&20.3&42.6&81.1\\
0.45&28.0&16.2&-2.0&30.0&14.6&40.8&87.4\\
0.50&25.5&14.8&-5.0&29.8&15.3&40.4&89.7\\
\hline\hline
\end{tabular}
\end{table}

\begin{table}[h]
\centering
\caption{Selected covariance matrices, in mb$^2$.}
\begin{tabular}{c|rrr}
\hline\hline
$B_\Lambda=0.10$ MeV & breakup & $\Lambda\Sigma$ & absorption\\
\hline
breakup&3140.3&-490.7&-2695.2\\
$\Lambda\Sigma$&-490.7&1418.6&-627.9\\
absorption&-2695.2&-627.9&5581.4\\
\hline
$B_\Lambda=0.30$ MeV & breakup & $\Lambda\Sigma$ & absorption\\
\hline
breakup&680.9&-89.4&-586.3\\
$\Lambda\Sigma$&-89.4&1060.1&-828.3\\
absorption&-586.3&-828.3&2371.2\\
\hline
$B_\Lambda=0.50$ MeV & breakup & $\Lambda\Sigma$ & absorption\\
\hline
breakup&218.3&6.5&-257.3\\
$\Lambda\Sigma$&6.5&888.3&-794.8\\
absorption&-257.3&-794.8&1636.1\\
\hline\hline
\end{tabular}
\end{table}

The absorption channel accounts for approximately 80--90\% of the total variance, the $\Lambda\Sigma$ channel for roughly 10--20\%, and breakup has a small net covariance share. These percentages refer to uncertainty propagation, not to the central channel strengths.

\textit{Bayesian prior and boundary effect.} -- The posterior is $p(B_\Lambda|D)\propto p(D|B_\Lambda)p(B_\Lambda)$, with a uniform prior on $0.1\le B_\Lambda\le0.5$ MeV. The likelihood combines the experimental error and the interpolated theory covariance. Near the endpoints, the prior truncates one side of the likelihood, reducing the inferred credible width. Repeating the inference with a broader interval would raise the endpoint widths and flatten the broad maximum, but it would not remove the large information content set locally by $\delta B_\Lambda\sim\delta\sigma_I/|d\sigma_I/dB_\Lambda|$.

\textit{Coulomb breakup.} -- The long-range Coulomb eikonal phase cannot be inserted naively because it produces the familiar divergence in first-order breakup amplitudes. For the light ${}^{12}$C target used here, a separate finite Coulomb-breakup calculation gives 8.57 mb at 1.5 GeV/nucleon, much smaller than both the nuclear interaction cross section and its propagated uncertainty. Coulomb breakup is therefore omitted from the quoted $\sigma_I$. The corresponding value for ${}^{208}$Pb is about 1.39 b; heavy targets require an explicit finite Coulomb treatment and lie outside the present scope.

\end{document}